\documentstyle[sprocl,epsfig]{article}

\def\NIM{\em Nucl. Instrum. Methods}

\def\NPB{{\em Nucl. Phys.} B}
\def\PLB{{\em Phys. Lett.}  B}
\def\PRL{\em Phys. Rev. Lett.}
\def\PRD{{\em Phys. Rev.} D}

\def\mjjtautau{M_{jj\tauptaum}}

\def\be{\begin{equation}}
\def\ee{\end{equation}}
\def\bea{\begin{eqnarray}}
\def\eea{\end{eqnarray}}
\def\tauptaum{\tau^+\tau^-}

\def\h{h}
\def\mh{m_{h}}

\def\hbar{\overline h}

\def\mx{M_X}

\def\h{h}
\def\mh{m_{\h}}

\def\tauptaum{\tau^+\tau^-}

\def\lsim{\mathrel{\raise.3ex\hbox{$<$\kern-.75em\lower1ex\hbox{$\sim$}}}}
\def\gsim{\mathrel{\raise.3ex\hbox{$>$\kern-.75em\lower1ex\hbox{$\sim$}}}}
\def\ifmath#1{\relax\ifmmode #1\else $#1$\fi}

\def\susy{{\rm SUSY}}

\def\mhh{m_{\hh}}
\def\mhl{m_{\hl}}
\def\hh{h_H}
\def\hl{h_L}
\def\ma{m_{a}}

\def\mtau{m_\tau}
\def\mb{m_b}

\def\mx{M_X}

\def\MPL #1 #2 #3 {{\sl Mod.~Phys.~Lett.}~{\bf#1} (#3) #2}
\def\NPB #1 #2 #3 {{\sl Nucl.~Phys.}~{\bf #1} (#3) #2}
\def\PLB #1 #2 #3 {{\sl Phys.~Lett.}~{\bf #1} (#3) #2}
\def\PR #1 #2 #3 {{\sl Phys.~Rep.}~{\bf#1} (#3) #2}
\def\PRD #1 #2 #3 {{\sl Phys.~Rev.}~{\bf #1} (#3) #2}
\def\PRL #1 #2 #3 {{\sl Phys.~Rev.~Lett.}~{\bf#1} (#3) #2}
\def\RMP #1 #2 #3 {{\sl Rev.~Mod.~Phys.}~{\bf#1} (#3) #2}
\def\ZPC #1 #2 #3 {{\sl Z.~Phys.}~{\bf #1} (#3) #2}
\def\IJMP #1 #2 #3 {{\sl Int.~J.~Mod.~Phys.}~{\bf#1} (#3) #2}
\def\NIM #1 #2 #3 {{\sl Nucl.~Inst.~and~Meth.}~{\bf#1} {#3} #2}
\def\br{B}
\def\tauptaum{\tau^+\tau^-}

\def\gam{\gamma}

\def\anti{\overline}
\def\epem{e^+e^-}
\def\mupmum{\mu^+\mu^-}

\def\mupmum{\mu^+\mu^-}

\def\rts{\sqrt s}

\def\anti{\overline}

\def\gev{~{\rm GeV}}

\def\mb{m_b}

\newcommand{\nc}{\newcommand}
\nc{\beq}{\begin{equation}}   \nc{\eeq}{\end{equation}}
\nc{\baa}{\begin{array}}      \nc{\eaa}{\end{array}}
\nc{\bit}{\begin{itemize}}    \nc{\eit}{\end{itemize}}
\nc{\ben}{\begin{enumerate}}  \nc{\een}{\end{enumerate}}
\nc{\bce}{\begin{center}}     \nc{\ece}{\end{center}}
\def\beqa{\begin{eqnarray}}
\def\eeqa{\end{eqnarray}}
\def\bed{\begin{description}}
\def\eed{\end{description}}

\def\ee{e^+e^-}

\begin{document}

\title{NMSSM HIGGS DETECTION: LHC, LC, $\gam$C
COLLIDER COMPLEMENTARITY AND HIGGS-TO-HIGGS DECAYS~\footnote{To appear
  in the Proceedings of the International Conference on Linear Colliders, Paris, April 19-23, 2004.}\vspace*{-.17in}}

\author{JOHN F. GUNION$^1$, MICHA{\L} SZLEPER$^2$}

\address{$^1$ Department of Physics, University of California at
  Davis, Davis CA 95616 \\ $^2$ Department of Physics,
Northwestern University, Evanston, IL 60208}


\maketitle\abstracts{\vspace*{-.1in}
We discuss the importance of being able to 
detect Higgs-to-Higgs-pair decays in
the context of the Next-to-Minimal Supersymmetric Model (NMSSM) and
demonstrate the excellent capabilities of a photon
collider for this purpose.
\vspace*{-.1in}}
  
%***********************************************************************
\section{Introduction}
%***********************************************************************
The Minimal Supersymmetric Model (MSSM) has been
the canonical benchmark for \susy\ studies for many years.
However, LEP constraints demand parameter choices for
which the fine-tuning and little hierarchy problems
have become quite an issue. And, of course, there
is still no truly attractive explanation of the $\mu$
parameter of the MSSM.  
The NMSSM relaxes the fine-tuning and hierarchy problems
and provides a simple explanation for an electroweak-scale
value for $\mu$. As such, it deserves at least as much
attention as the MSSM. 

It has been demonstrated  \cite{Ellwanger:2004gz}
(see also references therein) that it may be very difficult
to guarantee discovery of even one of the NMSSM
Higgs bosons at the LHC (for parameter choices not
already ruled out by LEP) because of the possibility
of Higgs-to-Higgs decays. In particular, there are many choices for NMSSM
parameters for which all the standard LHC production/decay channels 
used for guaranteeing discovery of at least one MSSM
Higgs boson [these include: 1) $g g \to h/a \to \gamma \gamma$;
2) associated $W h/a$ or $t \bar{t} h/a$ production with 
$\gamma \gamma\ell^{\pm}$ in the final state;
3) associated $t \bar{t} h/a$ production with $h/a \to b \bar{b}$;
4) associated $b \bar{b} h/a$ production with $h/a \to \tau^+\tau^-$;
5) $g g \to h \to Z Z^{(*)} \to$ 4 leptons;
6) $g g \to h \to W W^{(*)} \to \ell^+ \ell^- \nu \bar{\nu}$;
7) $W W \to h \to \tau^+ \tau^-$;
8) $W W \to h\to W W^{(*)}$] fail by virtue of the fact that the only
strongly produced relatively light Higgs boson (denoted $\hh$) decays with
branching fraction near one to two
other Higgs bosons, $\hl\hl$ (most frequently $\hl=a$, the lightest
pseudoscalar). For $\mhl>2\mb$, each $\hl$  decays
to $b\anti b$ and $\tauptaum$ (in roughly 0.9:0.1 ratio).
For $2\mtau<\mhl<2\mb$, the $\tauptaum$ mode is dominant, with
$jj=c\anti c +s\anti s +gg$ making up the rest. In all such cases,
the $\hh$ has relatively SM-like couplings, in particular to $WW$, and mass
in the $[75\gev,150\gev]$ range (with $\mhh\in[100\gev,120\gev]$ being
particularly probable).

The LHC detection channel $WW\to \hh\to\hl\hl\to jj\tauptaum$ (where $j=b$
when $\mhl>2\mb$) was explored in \cite{Ellwanger:2004gz}, with the
conclusion that (after appropriate cuts) a signal could be seen
above the (dominant after cuts) $t\anti t$ background as a broad-bump
excess in the $\mjjtautau$ mass distribution 
at low $\mjjtautau$ above a rapidly falling (as $\mjjtautau$
decreases) $t\anti t$ continuum. While nominally large statistics are
obtained for this signal, it will be the only signal
(other than
the perturbativity of $WW\to WW$ scattering)
for the presence of any of the NMSSM Higgs bosons in these cases.  
If this kind of scenario is nature's choice, a proper
study of the NMSSM Higgs bosons will probably only be possible at
a linear $\epem$ collider, or as discussed here, a $\gam\gam$
collider.

At a $\rts>350\gev$ $\epem$ collider, one would simply use $\epem\to ZX$ to detect
the $\hh$ Higgs bump in the missing mass $\mx$. Once detected, it is an
easy task to explore the $\hh$ decays. In particular, if $\hl\hl$
is the dominant decay mode, one could look also at the branching
fractions of the $\hl$ to different modes ($b\anti b$, vs. $\tauptaum$
vs. $gg+c\anti c +s\anti s$) and check that they are in
the ratio expected for a light Higgs boson (with no or suppressed $WW$
coupling) with the observed $\mhl$.

However, it is not clear when such an $\epem$ collider will be built.
It is possible that a low-energy $\gam\gam$ collider based on
the use of a pair of CLIC modules \cite{Asner:2001vh} 
might be the first collider able to
produce the $\hh$ in a sufficiently clean experimental environment
that detection of $\hh\to\hl\hl$ in a variety of the $\hl$ decay channels
would be possible. In what follows, we show that this potential is
fully realized. Thus, if \susy\ signals consistent with a model like
the NMSSM are seen at the LHC, but there is no or only a very weak
signal for any of the Higgs bosons, then a low-energy $\gam\gam$
collider will become a priority, especially 
if it can be built before (but also even if only 
as a facility at) a full-scale $\epem$ collider. 
\vspace*{-.1in}

\section{The $\gam\gam$ Collider}

We review results for the $\hh=h$, $\hl=a$ cases that would
be typical of the NMSSM. For most parameter choices, $m_a>2\mb$
and the $b\anti b$ and $\tauptaum$ branching ratios would be of order
$0.85-0.9$ and $0.06$ to $0.1$, respectively.
Results presented assume that the primary $h$ has
SM-like $\gam\gam$ production rate.  This is typical of the NMSSM
cases that would escape LHC detection in traditional modes. 
Sensitivity at a 
  $\gam\gam$ collider  for a Higgs boson that
decays primarily to $aa$ but does not have SM-like $\gam\gam$
coupling can be obtained by rescaling.\def\underl{\underline}
The most important signal channels are
\ben\itemsep=-.05in
\item $\gam\gam\to h\to aa\to b\anti b b\anti b$, $\leftarrow$ \underl{done}
\item $\gam\gam\to h\to aa\to b\anti b \tauptaum$, $\leftarrow$ \underl{done}
\item $\gam\gam\to h\to aa\to \tauptaum\tauptaum$. $\leftarrow$ \underl{not yet done} 
\een
The important backgrounds are:
 $\gam\gam\to b\anti b b\anti b$, $b\anti b c\anti c$, $c\anti c c\anti c$;
 $\gam\gam\to b\anti b \tauptaum$, $c\anti c \tauptaum$; and 
$\gam\gam\to \tauptaum\tauptaum$, depending upon the final state above.

For the signal, we employed an appropriately kluged version of
Pythia 6.158 \cite{Sjostrand:2003wg} interfaced with CAIN \cite{cain}
for correct $\gam\gam$ luminosity spectra. 
For the background, we employed WHIZARD 1.24 \cite{whizard}. The
cross sections for 4-fermion processes were cross checked with
theoretical computations for $\gam\gam\to
\epem\epem$, $\mupmum\mupmum$, $\epem\mupmum$ \cite{Carimalo:2000nu}. The
cross sections for $\gam\gam\to b\anti b$ and $\gam\gam\to c\anti c$ 
we obtain are consistent with Pythia, including beam polarization
effects. We find that it is easy to develop tagging and cut procedures
that effectively eliminate the backgrounds listed above. 

     \begin{figure}[t!] 
       \vspace*{-.75in}
     \begin{center}
\hspace*{-1.1in}
     \epsfig{file=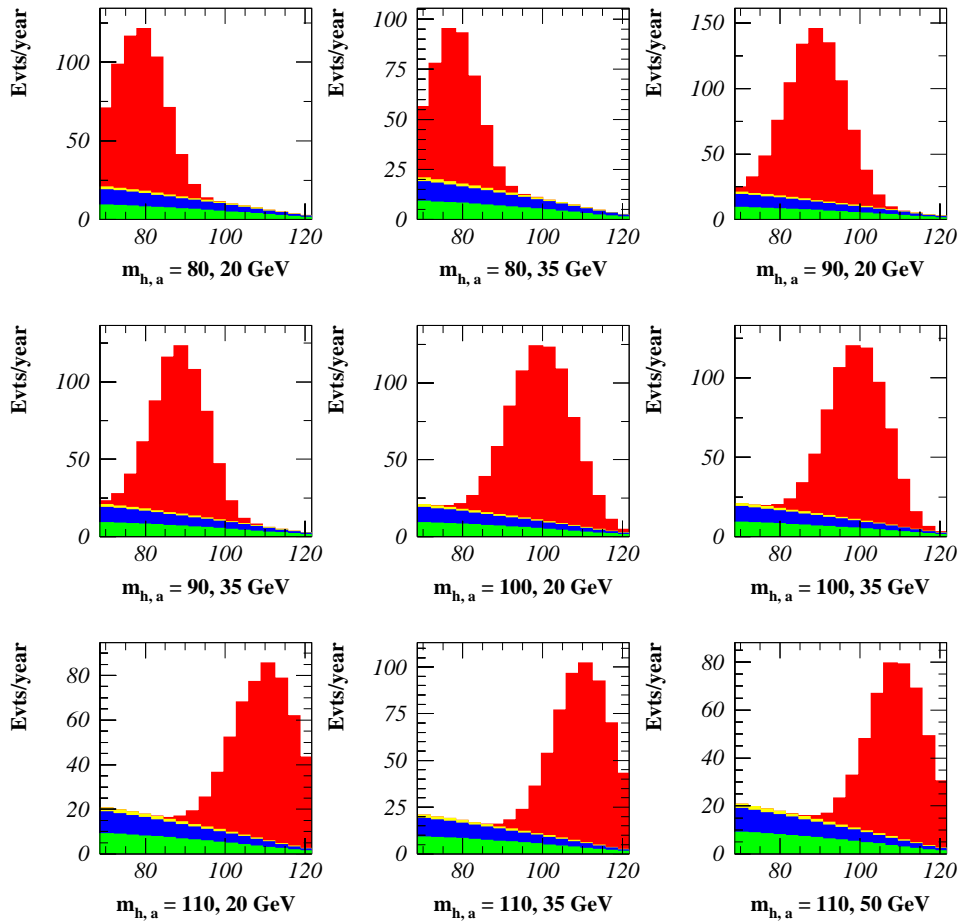,angle=90,width=20.5cm}
     \end{center}
     \vspace*{-.5in}
     \caption{Signal (big red peak) vs. backgrounds ($4b$ -- green; $2b
      2c$ -- blue; $4c$ -- yellow) for $\gam\gam\to h\to
       aa\to 4b$ for various $m_h$ and $m_a$ choices. $E_e=75\gev$
       broad-spectrum results.}
\vspace*{-.25in} 
     \label{broad}
     \end{figure}

     \begin{figure}[t!] 
     \begin{center}
\hspace*{-.2in}
     \epsfig{file=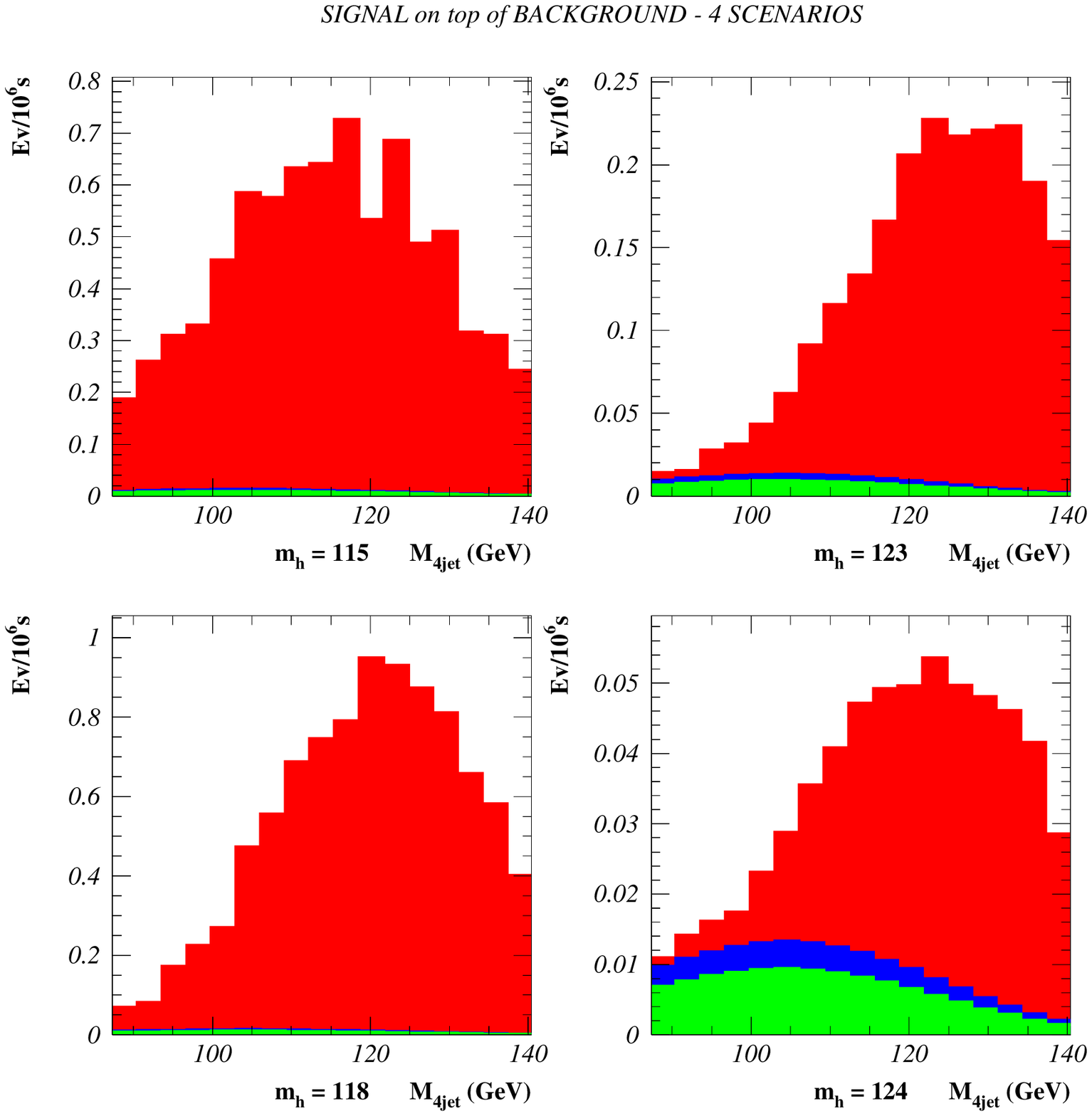,width=12cm}
     \end{center}
     \caption{Signal (big red peak) vs. $b\anti b \tauptaum$
and $c\anti c\tauptaum$ backgrounds ($2b2\tau$ -- green; $2c2\tau$ --
blue)
for $\gam\gam\to h\to
       aa\to 2b2\tau$ for $(\mh,\ma)=(115,56)$, $(123,35)$, $(118,41)$
and $(124,59)\gev$. $E_e=75\gev$ peaked-spectrum results.} 
     \label{peaked}
     \vspace*{-.2in}
     \end{figure}

 If there is very weak knowledge of the Higgs mass from the LHC, 
one would wish to explore the largest possible range of masses at
the $\gam\gam$ collider.  For nominal
CLIC single module, single beam energy of $E_e=75\gev$, one can adjust
the laser polarizations so that the back-scattered photons 
have an $E_{\gam\gam}$ spectrum which covers a broad
range up to $\mh\leq 115\gev$.  By increasing $E_e$ to $82\gev$
(as is apparently technically feasible) Higgs discovery in the broad
spectrum mode of operation would be
possible up to $\mh\leq 125\gev$.  To go still higher in $\mh$ (or to cover the
$\mh\in[115\gev,125\gev]$ region at $E_e=75\gev$) would require using
the laser polarization configuration that gives a $E_{\gam\gam}$
spectrum that is strongly peaked at $E_{\gam\gam}$ values just below
the upper limit (for example, the peak is close to
$E_{\gam\gam}\sim 115\gev$ for $E_e=75\gev$). Once the Higgs is
detected in $\gam\gam$ collisions, its mass will be determined rather
precisely; one would then employ the peaked spectrum
configuration and an $E_e$ such that the $E_{\gam\gam}$ peak
is centered at $\mh$.

A sample of our $E_e=75\gev$ broad-spectrum results
appears in fig.~\ref{broad}, where we show the signal in the $4b$
final state invariant mass spectrum
(after cuts and $b$-tagging) for a grid of choices:
$m_h=80$, $90$, $100$, and $110\gev$ and $m_a=20$, $35$, and $50\gev$,
yielding 9 cases for which $h\to aa$ is kinematically allowed.
In all these cases, the $a$ mass could be determined quite precisely
from the $2b$ mass spectra.
A sample of our $E_e=75\gev$ peaked-spectrum results for the $2b2\tau$
final state (where the $\tau$'s decay to a low-multiplicity ``jet'' +
neutrinos and the $2\tau$ mass is reconstructed by using
$p_T$ balancing and the assumption of collinearity between
visible and invisible momentum for each $\tau$ decay) appears in
fig.~\ref{peaked}. There we display the ``$4j$'' mass spectra in
the cases $(\mh,\ma)=(115,56)$, $(123,35)$, $(118,41)$
and $(124,59)\gev$; the signal peaks contain 78, 20, 92 and 4.5
events, respectively.  The small numbers in the 2nd and 4th cases
are simply a result of the fact that $\mh$ is significantly beyond the 
$E_{\gam\gam}\sim 115\gev$ peak (at $E_e=75\gev$); the $E_{\gam\gam}$
spectrum runs out just above $E_{\gam\gam}=125\gev$.
 (Better event rates would be obtained by running the CLIC
modules at $E_e=82\gev$, as possible.) In general, so long as
one is not severely event-rate limited, a rather accurate
determination of ${\br(a\to b\anti b)\over \br(a\to \tauptaum)}$
would be possible from the ratio of the $4b$ to $2b2\tau$ event
rates.

Besides $h\to aa$ decay cases, we have also explored 
cases in which $h\to a_1 a_2$ with $m_{a_1}\neq m_{a_2}$ (not relevant
for the difficult NMSSM cases, but possibly relevant in
the context of other models, including the CPX MSSM
\cite{Carena:2002bb} where $a_1$ and
$a_2$ would actually be CP-mixed states). There, one must repeat the analysis
for various choices of $m_{a_1}$ and $m_{a_2}$ and look for that
choice which maximizes the signal rate. In fact, we find that by
doing so not only is an excellent signal obtained, but
also it is possible to determine
the two masses with reasonable accuracy.

\section{Conclusion}
\vskip -.1in

Because of increasing constraints on the Higgs sector,
the MSSM can no longer be regarded as the most attractive
{\it and} simple supersymmetric model.  This honor belongs to the NMSSM, where
the addition of just one extra singlet superfield solves the
$\mu$-problem and greatly ameliorates
the now-apparent 
fine-tuning and little-hierarchy problems of the MSSM. The price
is the possible difficulty of detecting any of the NMSSM Higgs bosons
at the LHC. For a large part of NMSSM parameter space the
Higgs-to-Higgs-pair $\hh\to\hl\hl$ 
decays are important or dominant for the most
SM-like Higgs boson. When dominant, the only Higgs signal at the LHC
will be a broad-bump in the $jj\tauptaum$ signal. If not dominant,
other LHC signals may be present, but understanding the NMSSM Higgs
sector will still require direct observation of the $\hl\hl$ final
state. An $\epem$ linear collider is certainly capable of exploring
such decays using $\epem\to Z\hh$ production. Here, we have
summarized the very excellent ability of a two-CLIC-module based
(or other low-energy) $\gam\gam$ collider to provide
a detailed survey of  
$\gam\gam\to \hh\to\hl\hl$ production/decay in the most important $\hl\hl$
final state decay modes. In general, the $\gam\gam$ collider will be an
excellent facility and complement to the LHC 
for studying any $\hh$ with SM-like properties, regardless
of its dominant decay mode(s) ({\it cf.} \cite{Asner:2003hz}).

\section*{Acknowledgment}

We thank M. Velasco for suggestions and discussions. 

\section*{References}

\bibliography{nmssm_lcws04_paper.bib}

\end{document}